# Behaviour in social media for floods and heat waves in disaster response via Artificial Intelligence

**Víctor Ponce-López [1] \*, Catalina Spataru [2]**

[1] UCL Energy Institute; v.poncelopez@ucl.ac.uk

[2] UCL Energy Institute; c.spataru@ucl.ac.uk

**\*** Correspondence : v.poncelopez@ucl.ac.uk ; c.spataru@ucl.ac.uk Tel.: +44 (0) 20 7679 2000

**ABSTRACT**

This paper analyses social media data in multiple disaster-related collections of floods and heat waves in the UK. The proposed method uses machine learning classifiers based on deep bidirectional neural networks trained on benchmark datasets of disaster responses and extreme events. The resulting models are applied to perform sentiment and qualitative analysis of inferred topics in text data. We further analyse a set of behavioural indicators and match them with climate variables via decoding synoptical records to analyse thermal comfort. We highlight the advantages of aligning behavioural indicators along with climate variables to provide with additional valuable information to be considered especially in different phases of a disaster and applicable to extreme weather periods. The positiveness of messages is around 8% for disaster, 1% for disaster and medical response, 7% for disaster and humanitarian related messages. This shows the reliability of such data for our case studies. We show the transferability of this approach to be applied to any social media data collection.

## 1 Introduction

In the context of crisis and disaster management, advisory documents are becoming available to explore how different regions address crisis and disaster events from their governmental institutions. For instance, the Centers for Disease Control and Prevention [1] from the U.S. Department of Health and Human Services provide with guidance in terms of preparedness and promoting safe practices about how and when to inform about hurricanes, flooding, and similar disasters. Similarly, The UK Department for Environment Food and Rural Affairs developed a summary of recommendations in response to the multi-agency flood plan specially in terms of resilience and resource funding. In the case of heat waves, there is a visual electronic source from the U.S. National Weather Service (Reference 1 in S1 Supporting Information) which provided with extensive advice in heat safety for summer 2020. More generally, a review article in global health from [2] described the main natural or mankind events which caused the serious challenges for public health, and a guide in emergencies was published by the John Hopkins and Red Cross Red Crescent [3].

Over time, social media grow rapidly and considered an important source of additional knowledge to be integrated in the analysis for responding to crisis and disaster. However, the use of social media data is challenging in many different aspects, which have to do with filtering valuable and reliable information in nearly real-time and detect hot spots where to action when an event occurs or is about to happen. Besides, novel works in sentiment analysis and engagement prediction have shown to support the analysis by the combining social sensing with multiple machine learning approaches.

This paper presents qualitative analysis of our methodology framework, designed to combine, and apply artificial intelligence modules which provide novel insights in the context of disaster response. We base on a filtering approach to subsequently apply topic inference models on data collections for heat waves and floods. These techniques are used both to identify relevant categories and to provide a semi-automatic approach to feedback the system with relevant messages from key user accounts. The filtering framework approach consist of elements to extract raw messages from social media data and to detect those belonging to relevant categories through multiple deep learning classifiers. The models for these classifiers learn main disaster-related categories via fine-tuning deep bidirectional Transformer neural network models, which are shown to boost performance in benchmark datasets of disaster response and extreme events. Then, we compute hand-crafted features modelled via LDA method to infer frequent disaster-related terms, and hence to statistically identify relevant topics. Lastly, we perform sentiment analysis and identification of behaviour indicators in our data collections, along with a decoding process of climate variables for the joint analysis of thermal comfort.

## 2 State-of-the-art of Social Media Data (SMD) Analysis in Disasters Research

The U.S. Department of Homeland Security [4] proposed innovative uses of social media for emergency management. The book from [5] also offers a detailed description of risks and global concerns of the population data, and tools and methods to address these complex problems in the context of disasters. A scientific review from [6] bases on the use of social media in emergency situations.

In the global perspective of natural hazards and disaster relief, many works emerged to use SMD for their understanding and assessment [7], detecting greater risks through patterns discovery [8], or deployed technologies and tools for data mining [9]. These tools and platforms have been deeply analysed along with their methodological considerations as important data sources of information to understand public health issues [10]. In this line, one can find a number of taxonomies for data analytics techniques which are developed along with social sensing and location-based systems [11], as well as reviews of platforms and crowdsourcing tools for disaster management [12].

In the context of extreme floods and heat waves, the vulnerability of coastal areas has been perceived in several cases studies; [13] and [14]. Parallel contributions to address this issue were in line of generating more structured datasets from historical weather records for coastal flooding [15], but also from general trends in flooding [16]. Besides, high-resolution population data served to improve estimations of global flood exposure [17]. As social media became more popular as a preferred communication channel, the need of effective informative messaging in emergencies also became challenging and it was addressed through terse messages in the case study from [18]. Case studies such as the Red River Valley Flood Threat relied on Twitter microblogging services, and they considered their implications significant to treat mass emergency events [19]. Twitter also took a similar role along with Facebook as organisational communication platforms in flood events in Northern Ireland [20], or Chennai flood towards disaster management [21], among others. Furthermore, high resolution social sensing of floods has also been shown to produce high-quality historical and real-time maps of floods when observing these type of natural hazards [22]. The fusion of social sensing with remote sensing has recently proved to derive informed flood extent maps via deep learning methods [23]. On the other hand, climate change has motivated to mining social media for identifying and tracking the impact of heat waves [24], crowdsensing [25], and their response from themes related to air conditioning, cooling center, dehydration, electrical outage, energy assistance, and heat [26].

At this point, data acquisition and processing remain crucial tasks in almost any analytic tool for SMD. Precisely, in the context of disaster response there is an added need to capture real-time data, which was analysed from the humanitarian logistics perspective in the research work from [27]. Big data tools take a key role in this process, and there is a variety of complex tools available for real-time acquisition such as [28]. Real-time stream analytics are widely used for a diverse of tasks including retrieval, topic classification, or clustering [29]. Natural language processing of SMD poses multiple challenges such as learning information categories, which require considerable amounts of annotated data for supervised classification tasks [30]. Moreover, combining semantic information extraction via machine learning techniques (Latent Dirichlet Allocation) with spatial and temporal analysis demonstrated to improve disaster management procedures [31]. In these research areas, there has been a number of works leveraging the use of deep learning methods such as convolutional neural networks to understand crisis-related data [32] or extreme flooding events [33]. The idea of combinig recurrent or convolutional neural networks that include an encoder and a decoder as part of the same model was introduced by Google Brain and Google Research with the work from [34], and it derived to future extensions of deep bidirectional neural networks which we base for our first step of the analysis in our paper. Besides, semantic compositionality resulted in considerable advances in the sentiment analysis domain through recursive deep models [35] along with other machine learning techniques [36], which conducted to applications in SMD such as in the political domain [37]. The emerging evolution of application domains of SMD analysis brought the scientific community to address novel general aspects such as engagement prediction [38] or behavioural contagion [39], but also specific ones in the context of disasters [40, 41]. At this time, novel techniques for fusing artificial intelligence tecniques have emerged in the context of disasters [42].

In the following sections, we describe the design and application of our framework which contains key components from the literature for the analysis of disaster-related SMD from textual information using machine learning and natural language processing. These components involve deep bidirectional Transformer neural networks for message filtering, inference models, sentiment and behaviour analysis, and a proposed approach to match behavioural indicators with decoded climate variables to analyse thermal comfort.

## 3 Methodology

In this section, we describe the method we use to analyse SMD in the domain of disaster response to assess the quality of SMD to identify relevant messages from authorities, journalists, volunteers, or other entities. The schematic diagram of the proposed framework for the methodology of this analysis is shown in Figure 1.

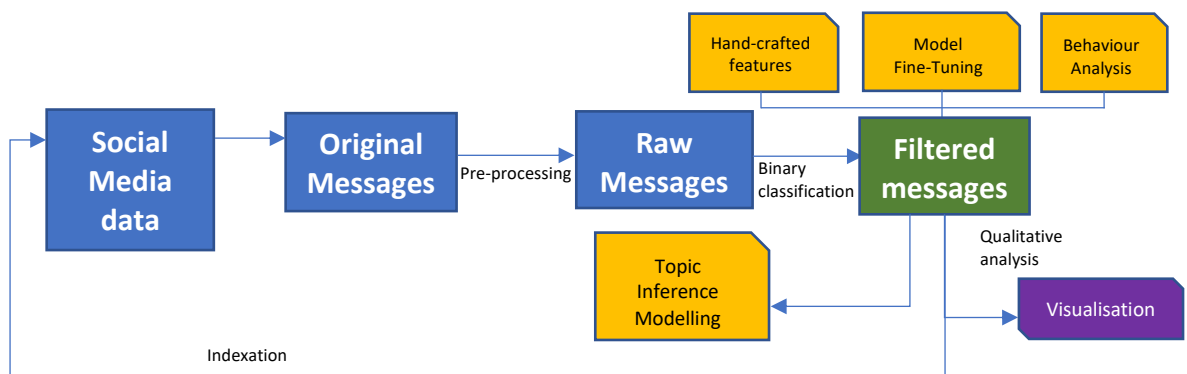

Figure 1: Schematic diagram of the proposed framework for the methodology of the analysis after pre-processing and filtering messages.

First, we pre-process the original messages from SMD collected that are related to disaster response to obtain raw messages.

Next, we apply the models from (Reference 15 in S1 Supporting Information) to filter certain messages related to disasters using binary classifiers. These models are based on deep neural networks to train distilled Bidirectional Encoder Representations (BERT) from [43] using the Transformers Python library from [44]. In particular, we aim to filter disaster-related messages that contain humanitarian and medical information which include any level of severity. Besides, we use a sentiment analysis classifier of another pretrained model from the same library to filter positive messages. The indexes obtained from the classification allows both to identify the filtered messages in the data corpus for the next steps and to recover any information from the original messages.

The following sections describe the processes for learning subsequent topics models from filtered messages, identifying behavioural indicators, and matching this information with climate variables of thermal comfort.

### 3.1 Topic Inference and Category Refinement

We describe the method we use to discover abstract topics that occur in our filtered data collections from their text using statistical modelling and, in particular, the Latent Dirichlet Allocation (LDA) method from [45]. This method allows to learn models from a corpus vocabulary to identify terms and subsequently classify messages into particular topics. Each topic consists of a set of words that are modelled as Dirichlet distributions.

The data pre-processing for this procedure involves tokenisation, removing short words and stopwords, lemmatisation of words into first person and verbs into present, and stemmisation of words into their root form. Once the data is pre-processed, we create a Bag-of-Words (BoW) dictionary containing the number of times a word appears in the set using Gensim [46]. We filter out tokens that appear either in less than 15 messages or in more than 0.5 messages as a fraction of the total corpus. Then, we keep only the first 100,000 most frequent tokens and we create a dictionary with the number of words and number of times those words appear. We use this dictionary to create LDA topic models from both Bag of Words (BoW) and Term Frequency Inverse

Document Frequency (TF-IDF) to explore the words occurring in each topic and their relative weight. Once a topic model is learned, we can use a test message to calculate the highest probability (classification accuracy) of the message to be part of the topic that the model assigned.

On the other hand, we subsequently use CrisisDPS from [47] to provide with a refined categorization to match the above topics. This system was developed to support humanitarian aid tasks automatically as an integration into disaster response workflows. The usage of this service takes input textual messages (e.g. our filtered messages from data collections) to determine for every message: the type of disaster; whether it is informative or not, and what type of humanitarian information it contains, using machine learning classifiers.

## 3.2 Text Analyisis and User's Behavioural Indicators

Our qualitative analysis focuses on the filtered messages from both data collections of floods and heat waves.

From the number of tools available for text analysis, we find the usage of ParallelDots framework and their API functionalities from (Reference 2 in S1 Supporting Information) a proper fit to automatically identify frequency and dominance of a variety of behavioural indicators in relevant messages. This tool utilizes models pretrained with machine learning techniques from [48] for the recognition of the well-known basic emotions from [49, 50]. Although further details about the machine learning models are not provided in the system's API documentation, we describe the list of five behavioural indicators and their confidence scores provided as category values which we consider in our analysis:

- **Sentiment:** negative, neutral, or positive. It describes the process of systematically identify, extract, quantify and study affective states and subjective information.

- **Emotions:** angry, excited, bored, fear, happy, or sad. It works at sentence level to extract words and phrases representing emotions or emotive states.

- **Intent**: feedback, marketing, news, query, or spam. It describes the type of information to identify genuine tweets and understand user's intentions.

- **Abuse**: abusive, hate speech, or neither. It focuses on filtering abusive content from a text corpus based on revealing connotation terms.

- **Sarcasm**: non-sarcastic, or sarcastic. It focuses on capturing components in a text block that aim to mock or convey contempt.

The system can be used either under a specific subscription or for free, having a limited number of daily hits that are reset over the month. Every hit is defined as the execution of a target query message to be analysed. To simplify the analysis, we use the free basic version using as input queries our reduced set of filtered messages resulting from the deep binary classifiers. Further details about the usage of the framework are provided along with the code documentation.

Besides the behavioural indicators, which we found of high value for our purpose in this research domain, the framework allows to recognise predefined and custom categories that are also relevant for our qualitative analysis. The predefined categories rely on the content taxonomy of IAB categories from (Reference 3 in S1 Supporting Information). This taxonomy is an industry standard containing levels of categories that are widely used in services like advertising, internet security and filtering appliances. Finally, the system allows the users to define custom categories for the messages to be classified, even though the authors do not provide with further specifications about this unsupervised or semi-supervised machine learning functionality.

In the context of disasters, we base these custom categories following advice from the National Governors Association [51] and the Economic Development Administration (Reference 4 in S1 Supporting Information), to define a set of categories related to the phases of a disaster:

- **Prevention** or Awareness: preventing human hazards from natural and manmade disasters.
- **Mitigation**: reducing loss of life and property by lessening the impact of disasters and emergencies.
- **Preparedness**: the continuous cycle of planning and organising.
- **Response**: the coordination and management of resources.
- **Recovery**: restoring critical functions to the community.

Due to the nature of messages in our data collections, we break down the recovery evaluation phase into needs and essentials for the purposes of resource allocation for our custom categorisation.

In the context of disaster response, governance of disaster risk reduction and resilience for sustainable development, we develop a list of potential stakeholders under all level of governance in the UK that directly or indirectly take part in one or several stages of disaster risk management from preparedness to recovery and mitigation. This includes national and sub-national public authorities, public authorities, category responders, sector resilience leads, non-governmental organisations, and civil society organisations. Per total a number of 212 stakeholders which have a Twitter account, most tweets were coming from MetOffice. followed by the community 'Climate Realists' and the U.K. Dept. of Health and Social Care.

We look for Twitter accounts for this list of users aiming to perform a semi-automatic identification of key messages and relevant usernames that might be first overlooked by the filtering approach. This allows to feedback the system by including new potential key messages from these users that are re-considered in the system with a reasonable effort of manual user inputs. By identifying these missing governmental and non-governmental institutions and other accounts in social media, we complement the behavioural analysis along with additional interactions of users to new key messages in terms of their reactions, responses, and forwarding (retweets).

### 3.3 Climate Variables of Thermal Comfort

We describe the procedure developed to decodify weather data encoded by meteorological stations into synoptic observations (SYNOP records) freely provided by the U.S. National Oceanic and Atmospheric Administration (Reference 5 in S1 Supporting Information). The goals of this decodification process are two sided. First, to reveal climate variables that are not yet available in annual reports from official sources such as the UK Meteorological Office (Reference 6 in S1 Supporting Information). Secondly, to obtain a more detailed information directly from weather stations without having been yet processed to generate the summaries for global human-readable weather reports.

We base on the six basic factors defined by the Health and Safety Executive (Reference 7 in S1 Supporting Information) to retrieve climate variables. Climate variables are related to thermal comfort (Reference 8 in S1 Supporting Information) and it is described by the HSE as a person's state of mind in terms of whether they feel too hot or too cold. The level of thermal comfort is also influenced by a combination of environmental, personal, and work-related factors. From these six basic factors, we aim to decode SYNOP weather data records for extracting the following human-readable climate variables and measurable units:

- Wind Speed (Knot (KT) converted to km/h).
- Maximum temperature (ºC).
- Average temperature (ºC).
- Relative Humidity.
- Precipitation (mm).
- Pressure (hPa).

Therefore, in Algorithm 1 we present the pseudocode of this decoding approach. First, SYNOP weather records are collected via queries in OGIMET system from (Reference 9 in S1 Supporting Information), then the decoding functions are based on (Reference 10 in S1 Supporting Information), which consists of a combination of scripting language functions in PERL and XML parsing for decoding a variety of reports, including synoptic observations. Further information about this procedure and form of SYNOP messages are provided along with the code documentation.

| | | |
|---|---|---|
| **Algorithm 1: Pseudocode for Decoding SYNOP Weather Records** | | |
| **Input** | : | Encoded SYNOP Document, $D$ |
| **Output** | : | Decoded data frame of climate variables, $DF$ |
| Step 1 | : | Read SYNOP document, $D$ |
| Step 2 | : | Initialize empty data frame, $DF \leftarrow DataFrame()$ |
| Step 3 | : | Initialize empty lists for each climate variable 'wind speed', 'max temperature', 'average air temperature', 'relative humidity', 'rainfall', and 'pressure'. $$CV \leftarrow \{ws, mt, at, rh, rf, pr \leftarrow list(), list(), list(), list(), list(), list()\}$$ |
| Step 4 | : | **for** each decoded message $m$ **in** $D$ **do** |
| Step 5 | : | Apply weather report PERL parser to generate decoded XML, $$doc \leftarrow metaf2xml(m)$$ |
| Step 6 | : | Get XML elements by decoded tag names for each target climate variable and append their values in the lists, $$CV \leftarrow \{ws \leftarrow getWSvalue(doc['wind']['speed']),$$ $$mt \leftarrow getMTvalue(doc['temp']['max']),$$ $$at \leftarrow getATvalue(doc['air']['temp']),$$ $$rh \leftarrow getRHvalue(doc['relHumid']),$$ $$rf \leftarrow getRFvalue(doc['precipitation']['precipAmount']),$$ $$pr \leftarrow getPRvalue(doc['stationPressure']['pressure']) \}$$ |
| Step 7 | : | **end for** |
| Step 8 | : | Dump set of values for every climate variable into a data frame, $$DF \leftarrow \{CV[var] \text{ for } var \text{ in } CV \text{ targets}\}$$ |
| Step 9 | : | **return** $DF$ |

We perform climate variable extraction using the steps described above on our target data collections described in the following section about data description. Extracting climate variables within the exact same periods of time as our SMD collections allows us to synchronise them with the previous behavioural indicators to provide combined insights for thermal comfort.

# 4 Data Collection

## 4.1 Twitter dataset

Tweets were collected during the year 2020 from Twitter SMD using the AIDR tool [28] and using the search terms highlighted in Table 1: Description of SMD collections related to disasters collected in 2020 for UKTable 1. The AIDR tool already incorporates a top-level filter by the search string passed to the Twitter API (Reference 11 in S1 Supporting Information). The filters considered in AIDR tool include crisis-specific labels [52] as a solution for domain adaptation.

All data were collected according to Twitter's terms of service and privacy conditions for a number of different events, including floods and heat waves. The different data collections are described in Table 1, including their periods of time and the main of parameters included in the AIDR system settings.

TABLE 1: DESCRIPTION OF SMD COLLECTIONS RELATED TO DISASTERS COLLECTED IN 2020 FOR UK

| Type of Event | Location | Period of Time | No # Tweets | List of Search Keywords |
|---|---|---|---|---|
| Floods | UK | Nov 2020 | 27K | UK flood, UK rainfall, Britain flood, Britain weather, extreme rainfall, London weather, London flood, flood warning, severe flood, severe rainfall, water hazard, flood hazard, hazard UK, flood day, flood alert |
| Heatwaves | UK | Jun-Jul. '20 Jul-Aug. '20 Aug-Sep. '20 | 602.5K | UK heatwave, UK hot weather, Britain weather, Britain heatwave, London weather, London heatwave, hottest day UK, Britain hottest day, summer hottest day, +30C, extreme heat, heatwave thresholds |

We present our general pre-processing methodology of our SMD collected for text standardization into raw input text for the machine learning classifiers.

For simplicity, in our qualitative analysis for this pilot study, we focus on the data collected in the UK during 2020 for floods and heat waves. These two first data collections provide us with a considerable amount of data both to feed the deep learning classifiers at the filtering stage, and to subsequently apply machine learning classifiers for obtaining the main insights of our overall analysis methodology described previously.

Something on validation

**4.2 Text Pre-Processing**

As mentioned first in the methodology section, we use the methodology from (Reference 15 in S1 Supporting Information) for text pre-processing from original messages in all our data collections. We use the Twitter pre-processing library from (Reference 12 in S1 Supporting Information) for cleaning mentions, hashtags, smileys, emojis, or reserved words such as RT. Then, we follow up the pre-processing with the BS4 library from (Reference 14 in S1 Supporting Information) to parse HTML URLs, and regular expression [53] operations in Python to remove anything that is not a letter of a number in all the messages. The result is a standardized raw text messages that are recognized both at the filtering step and the following stages of the presented methodology.

# 5 Qualitative Analysis

This section describes key results and insights from a qualitative analysis in our two main data collected for UK in 2020 with the AIDR tool for floods and heat waves. First, we explain our results related to frequent terms and relevant keywords for message filtering, identified topics, sentiment analysis, and category refinement. Next, we present results of text analysis and identification of several behavioural indicators. Finally, we show the procedure of matching those behavioural indicators with decoded climate variables related to thermal comfort.

## 5.1 Message Filtering, Topic Inference, and Sentiment Analysis

For the first part of our visualisation module, we use the Python Wordcloud library and their API functionalities from (Reference 13 in S1 Supporting Information) to generate visual interpretations of the most frequent terms and words sorted by number of occurrences.

These functionalities are applied to all raw text data obtained from pre-processing collections of UK floods and heat waves in 2020, which are shown in Figure 2 and Figure 3. The first noticeable aspect of our qualitative analysis is that the most frequent terms identified in these data collections reveal information related to weather warnings in the context of floods, as well as global climate warming and activism in the case of heat waves.

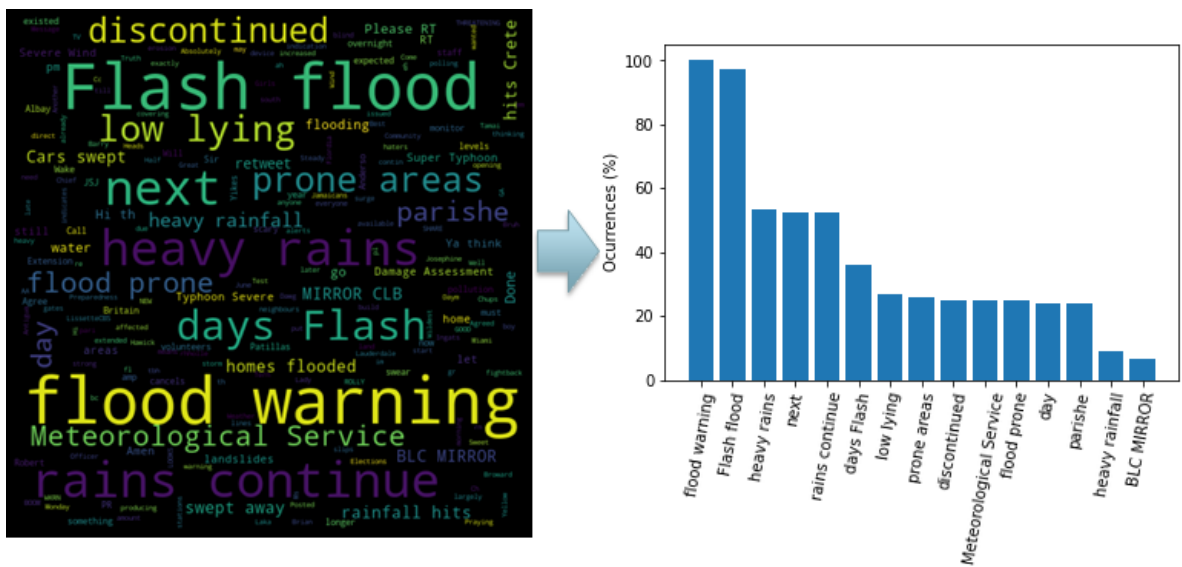

Figure 2: Word cloud chart and key terms plot in the UK Floods collection. 'Flood warning' and 'Flash flood' appear to be the most frequent words, followed respectively by the words 'heavy rains' and other terms related to the context of floods and weather.

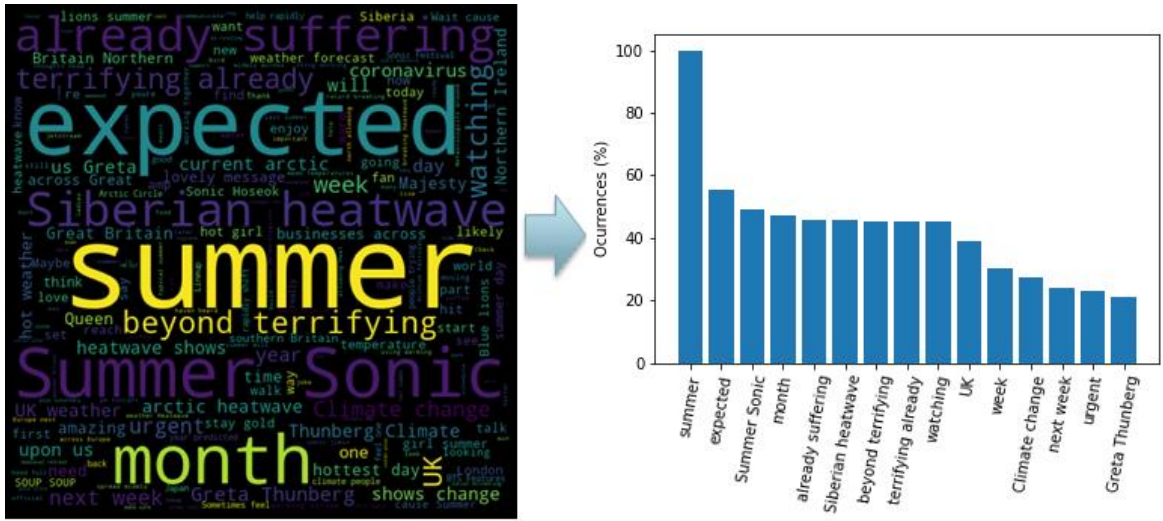

Figure 3: Word cloud chart and key terms plot in the UK Heatwaves collection. 'Summer' appears to be the most frequent word. Some other words related to activism and policies were identified in the context of global climate.

Moreover, the deep learning classifiers detected about 24K messages related to disasters out of the whole data for floods, which is a considerable amount with respect to the total number of collected messages. In the case of heat waves, about 91K out of a total subset of 100K messages from the data were considered to be related to disasters. The representation of the main categories related to disasters is shown in

Figure 4. For floods, we identified 1,098 medical-related messages and 275 related to humanitarian standards, whilst in the case of floods we identified 26,285 medical-related messages and 1,421 related to human standards.

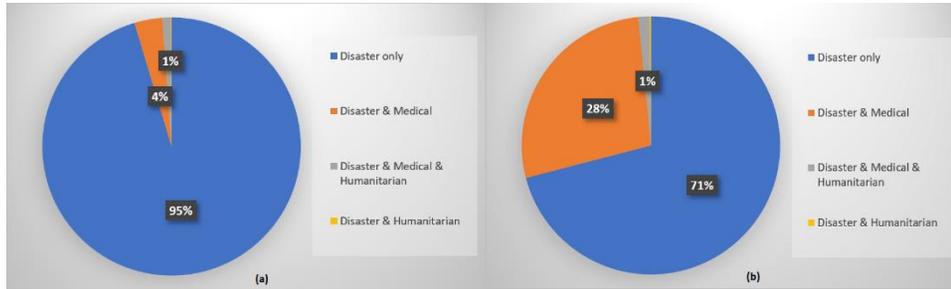

Figure 4: Percentage distribution of filtered messages for Floods (a) and Heatwaves (b) by main disaster categories: humanitarian standards and medical-related information.

Then, we applied LDA-based topic inference models to the messages likely related to disasters.

In the case of floods, the most frequent terms found by the system were 'warn', 'river, 'flash', 'water', 'hazard', and 'novemb', sorted respectively by descending relevance order. These terms gave us hints about potential topics related to informative weather messages or about warnings related to specific flooding events happening or going to happen. Other composite terms or keywords considered important were 'flood warning', 'flash flood', 'continues flood', and 'Flood Alert'. Many of those refer to informative messages from official local news and related sources. Besides, the dictionary of most frequent terms revealed words such as 'flood', 'detail', 'fair', 'form', ' 'fund', which appeared to be related to flood events their correlation with Covid-19 deaths or vulnerable areas, as well as to political decisions. Finally, the topic inference classifiers weighted again terms such as 'novemb', 'flash', 'warn', 'river', 'issue', 'till', which were related to official reports and information of the duration and location of the flooding events.

The sentiment analysis classifier identified 5,871 positive messages out of the 27,096 tweets.

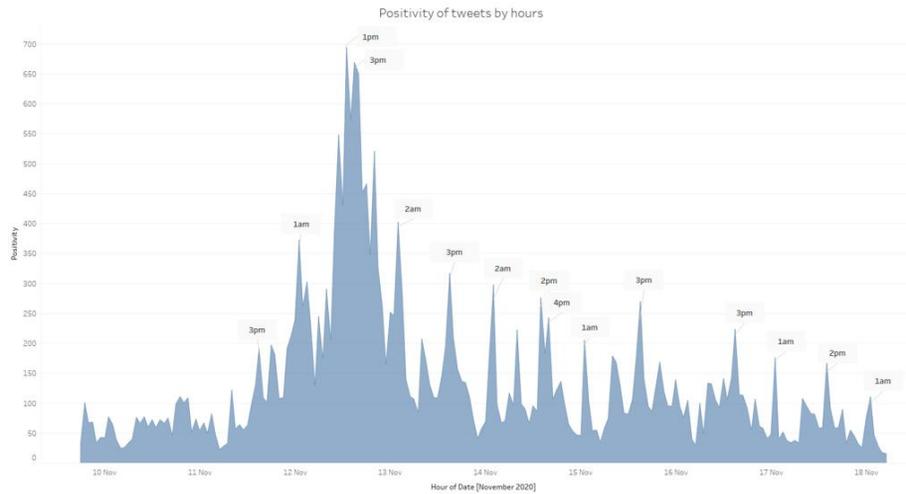

As shown in Figure *5*, the tweets were generally more positive around early afternoon and after midnight. An influencing factor of this was the increased activity at these hours, especially during the most rainy and active days, between 3pm on the 11th November and 2am on the 13th November, 2020. The overall hourly activity of users, however, appeared to be distributed differently along the whole period of 10 days of collected data. When we focus on the filtered messages, the positiveness of messages is around 17% (4,190 tweets), 27% (293 tweets) and 16% (45 tweets) for the disaster, disaster and medical; disaster and humanitarian related messages, respectively. Although many messages are from official sources, these low percentages of positiveness suggest that there was a considerable number of messages with negative connotations and a high influence of messages coming from other external sources. To dig into this issue, these signals indicate a potential need of increasing the amount of positive informative messages coming from official sources to alleviate the negative effects during these events.

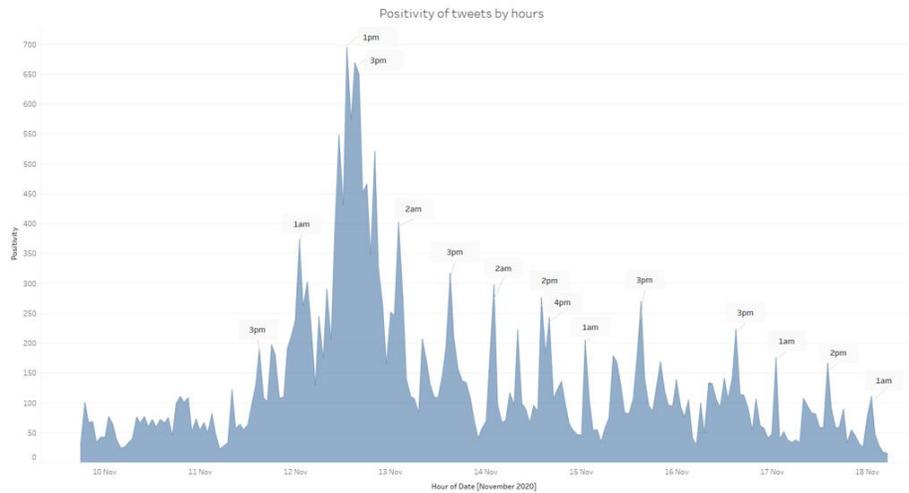

Figure 5: Sentiment analysis in the UK Floods data collection.

For heat waves, the most frequent terms found in the collection subset were 'heat', 'extrem', 'forget', 'year', 'record', 'weather', and 'drought', sorted respectively by descending relevant order. These terms gave us again hints about potential topics related to informative weather messages, but also about climate conditions and warnings related to events. Other composite terms or keywords considered important were 'fire fighter', 'extreme heat', 'climate change', 'heat ray', or 'threat'. Many of those refer to retweets from scientists about warnings related to extreme weather conditions increasing threats to property and life. Besides, the dictionary of most frequent terms revealed words such as 'decid', 'money', 'polic', 'fact', 'figther', 'fund', which appear to be related to political decisions or requests from activists related to the disaster response. Finally, the topic inference classifiers weights terms such as 'resourc', 'lack', 'fund', 'figther', 'fact', 'year', which relates to claiming for action in climate change and global warming.

The sentiment analysis classifier identified 31,467 positive messages out of the 99,967 tweets. As shown in Figure 6, the tweets were generally more positive in the evening with a few exceptions during some early morning and afternoon. An influencing factor of this was the increased activity in the evening, especially on the 26$^{th}$ August. The overall hourly activity of users, however, appeared to be distributed differently along this period of time. When we focus on the filtered messages, the positiveness of messages remains low around 8% (7,415 tweets), 1% (307 tweets) and 7% (107 tweets) for the disaster, disaster and medical, disaster and humanitarian related messages, respectively. Again, this low positivity suggests that many messages were coming from non-official sources having plenty of negative connotations and barely telling anything informative or valuable for the public. Again, to alleviate this effect along with an increased positivity of informative messages from official sources, another classification layer for informativeness would be helpful for the purpose of categorisation.

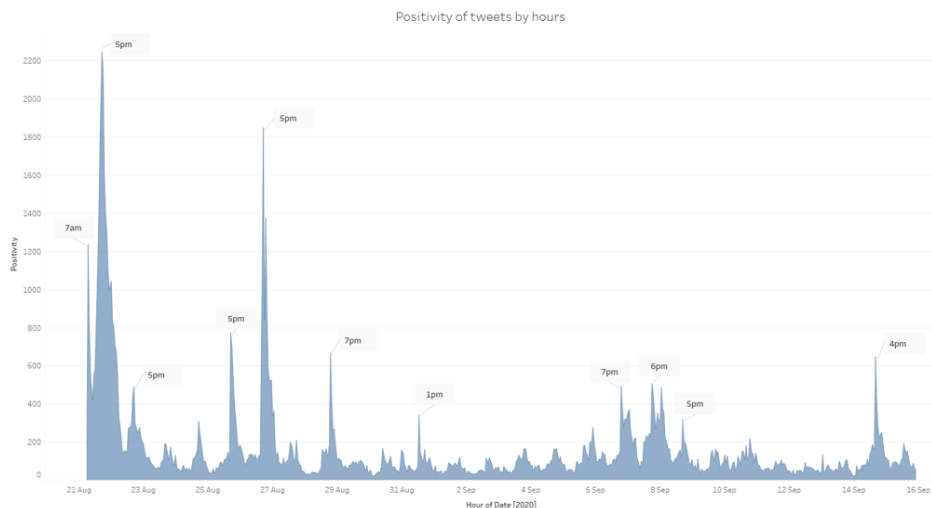
Figure 6: Sentiment analysis in the UK Heatwaves data collection.

From all set of positive filtered messages of the heat wave data collection, we end up with a total of 70 tweets that appear to contain information related to three main information categories: disasters, medical, and humanitarian standards. As a result of applying CrisisDPS using as input these filtered messages, the pie charts shown below in

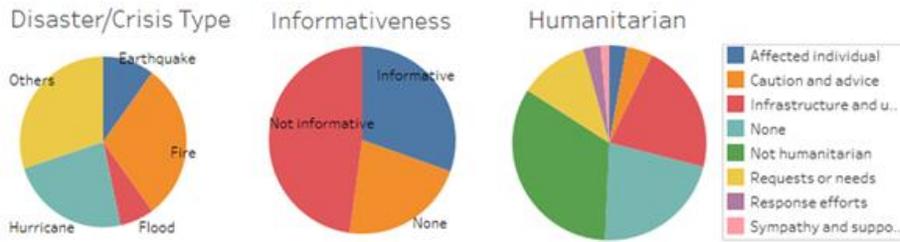

Figure *7* illustrate a summary of tweets falling into different types of disaster and crisis, as well as revealing their informativeness and type of humanitarian information. Still, we can notice a considerable amount of non-informative messages up to almost the half of them, and the type of disaster related to 'fire' which matches with the composite terms detected by our topic inference models. We highlight the usability of combining these methods from the identification of topics to a refined categorisation for the purpose of a focused disaster response.

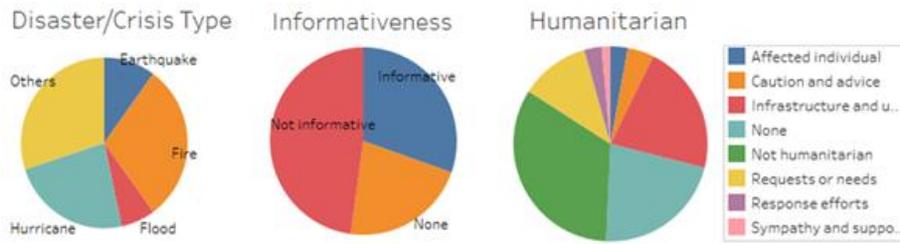

Figure 7: Sub-categorisation refinement for filtered messages from UK Heatwaves using CrisisDPS[47].

## 5.2 Text Behavioural Analysis

Here, we present the results of applying analysis of behavioural indicators described in the methodology, including both predefined and custom categories related to the phases of a disaster.

First, we conduct the analysis on the 36 resulting messages from filtering the UK floods data collection with our combination of deep binary classifiers presented previously. Figure 8 shows the results for the main IAB predefined categories (a) and custom categories broken down into detailed phases of a disaster (b). For the IAB taxonomy (a), we can notice the dominant categories are worldpost, green and impact, respectively, followed by entertainment and politics. Given the reduced number of filtered messages, we are able to show detailed results on this data collection for each message maintaining simplicity on the visualisation. The key filtered messages are related to reactions on the recent hurricanes, heavy winds, and storms, as well as news about flood warnings, alerts, and status of rivers. Most of these messages fall into those dominant categories as the most relevant categories in the context of disasters. Nevertheless, most messages from the categories 'politics' and 'entertainment' reveal recent actions or active discussions about political decisions and their consequences after the flooding event. There is almost any influence of the categories 'Crime', 'Religion', and 'Taste', for disaster-related messages. In the case of custom disaster categories (b), we can notice a predominance of the prior phases 'Preparedness', 'Mitigation', and 'Awareness', related to messages which are sent some days, hours, or minutes before a certain event happened. These messages are sent mostly from accounts related to news and meteorological services. For the posterior phases of a disaster, there is also a significant influence of 'Response' messages. However, in this case they are mostly related to reactions of people after the event when everything returned to normality. By contrast, the lesser confidence scores are obtained by the 'Need' category as part of the 'Recovery' phase. Although this is expected given the low severity of these events did not cause major trouble, there is a considerable presence of messages related to the category 'Eseential' which refer to resource allocation.

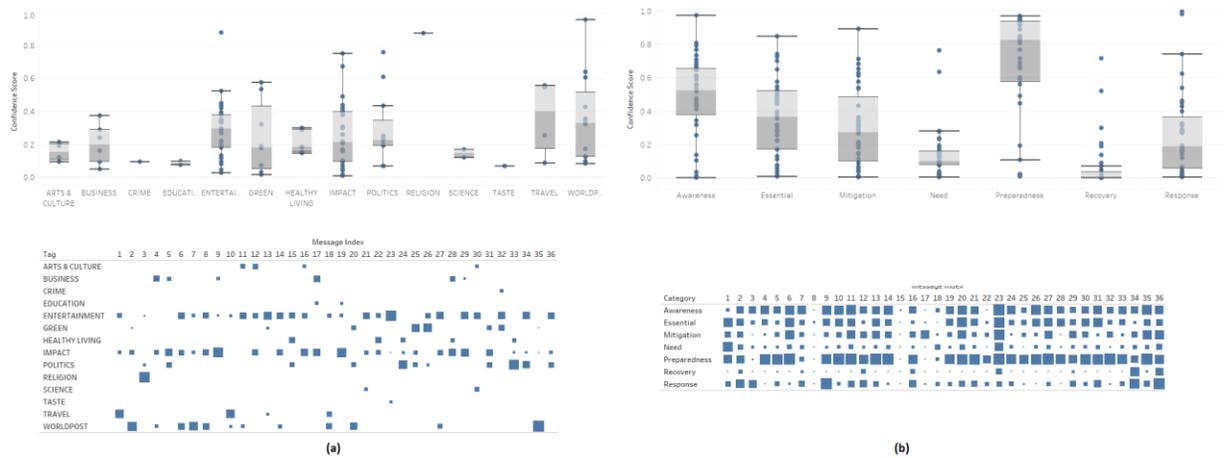

Figure 8: Distribution of averaged confidence scores and dominances of the main IAB categories for filtered messages of (a) Floods and (b) Heatwaves data collections in the UK. At the bottom, each column represents a filtered message assigned into each category, and the size of the squares represents the confidence score for that message in that category (the higher the score the bigger the square).

With regard to the behavioural indicators, in this case we conducted the analysis on filtered messages on both data collections. Figure 9 and Figure 10 show the results of the averaged distributions of dominance for each behavioural indicator. The distributions are quite similar on both datasets, and the behaviour indicator which appear to be the most equally distributed in its values of confidence scores is the 'sentiment'. The 'sarcasm' and 'abuse' indicators also behave similarly on both datasets, with a significant amount of up to almost 40% of messages considered sarcastic and a slightly increased number of abusive messages in the case of floods. Although both collections show, as expected, a predominance of news in their 'intent' indicator, we have seen an increased amount of filtered messages that are still considered as 'spam' in the heatwaves collection. This is likely to be caused by the higher number of total messages in this collection. We can also see an increased number of filtered messages classified as 'feedback' with respect to the floods collection. Lastly, the main relevant differences in terms of behavioural analysis when comparing both collections are in the 'emotion' indicator, where we can notice a significant increasement of angry messages in the floods collection, increased levels of fear and sad messages in the heat wave messages, but also in the exciting messages. Although these are surprising outcomes at a first glance, by digging into the heatwave collection we discover many messages related to global warming and environment, which are generally found to be those messages triggering the score levels on these three indicators.

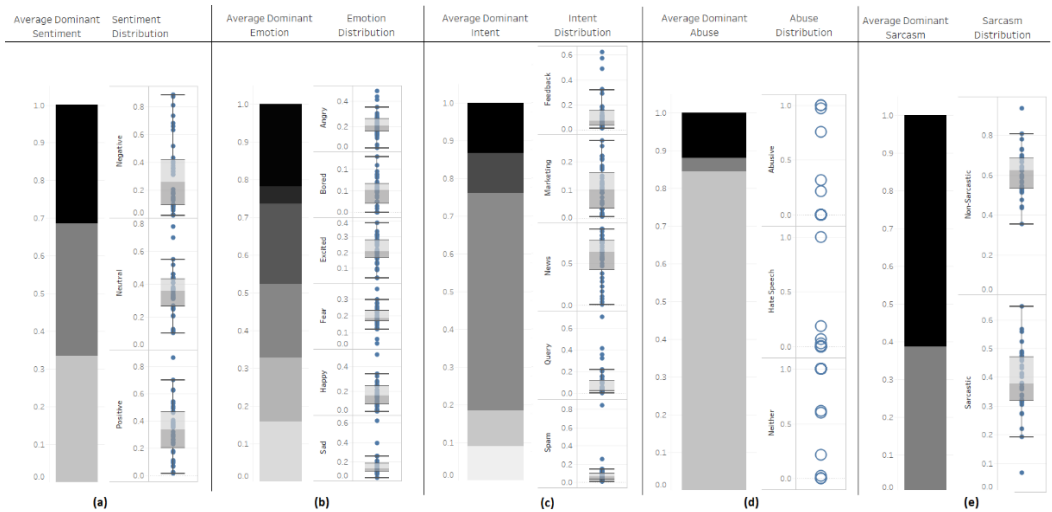

Figure 9: Average distributions of dominances for each behavioural indicators: sentiment (a), emotion (b), intent (c), abuse (d), and sarcasm (e), on filtered messages of the UK Floods collection

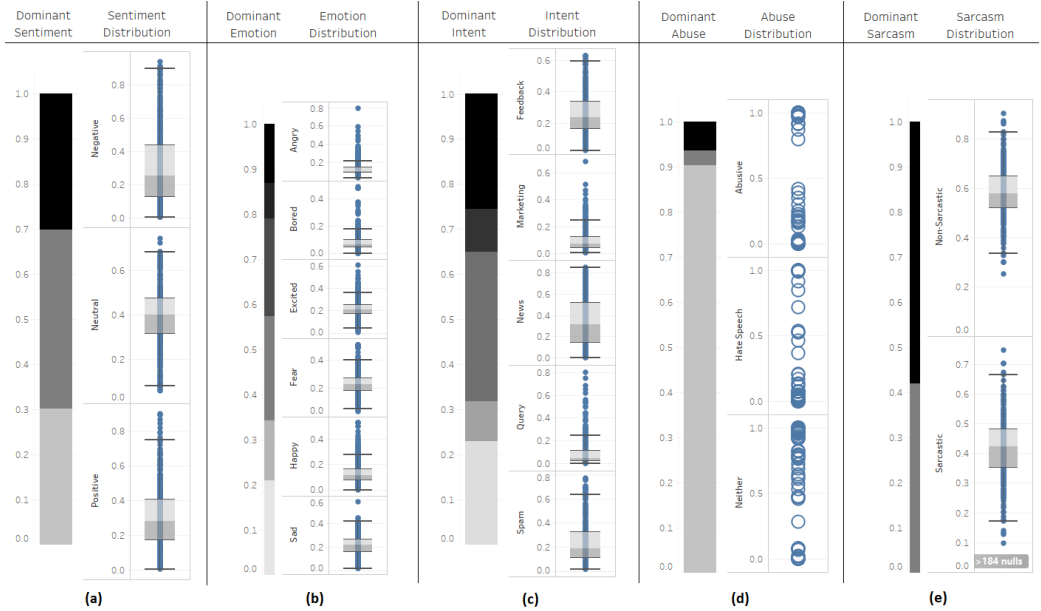

Figure 10: Average distributions of dominances for each behavioural indicators: sentiment (a), emotion (b), intent (c), abuse (d), and sarcasm (e), on filtered messages of the UK Heatwaves collection.

Moreover, we conducted a comparative analysis of key messages both from user accounts that are manually added from our list of well-known sources of information and those key messages that are detected via automatically filtering approach. From the list of known users described in the methodology, in the heatwave collection we found the highest occurrence of messages mostly coming mostly from national public authorities, such as official meteorological services, health and social care, and NHS; but also a considerable amount from well-known independent meteorologists and the scientific community in climate-related fields. These last two were almost at a similar level of occurrences than national public authorities.

On the other hand, the automatic filtering approach offers the possibility of discovering new potential candidates from the initial diversification of new information sources that become more popular in the network over time, and whose reliability is worth to be further evaluated to eventually include their knowledge, contributions, and value as a new relevant source of official information. This combined approach helped us to identigy major contributions of each independent process to eventually become a semi-automatic approach to feedback the whole system and better support responders at focusing on relevant information.

In conclusion, we found that a trade-off between sources of relevant information coming both from governmental institutions and those from non-governmental institutions can complement each other to enrich public knowledge and informativemess. Although this combination becomes crucial nowadays, it needs proper evaluation of reliability for new potential candidates to minimise desinformation, which is a field that remains out of scope for this paper.

**5.3 Matching Behaviour with Thermal Comfort**

We introduce a novel qualitative analysis with explanatory visualisations which combine textual analysis and behavioural indicators with the climate variables related to thermal comfort that we obtain with the decoding procedure presented in the end of the methodology section. We conduct this analysis in both entire collections of heat waves and floods in the UK during summer and winter in 2020, respectively.

Figure 11 illustrates a graphic where all plots are aligned to the same points in time. Starting from the top, the first and second plots show the count of positive messages and total activity messages, respectively, in the same way as explained first in this section for the sentiment analysis. Then, according to our methodology, we show values of average air temperatures, maximum temperatures, precipitation, pressure, relative humidity, and windspeed, respectively. Note the missing values for maximum temperatures and precipitation level; they match exactly with the capturing timescales of the weather stations, which are once per day in the case of maximum temperatures and a few times per day split into the same range of hours in the case of precipitation levels. Also note when some value is missing in any variable at a specific time, that point is removed for all variables to allow proper alignment and make the overall synchronisation consistent. Having pointed that, an aspect to highlight is the clear appreciation of increased levels of both precipitation and windspeed on the 11th November, followed by a considerably increased activity and positivity levels on the day after when these weather conditions ceased. Since there are no significant changes on the other climate variables, these increased levels potentially show a general improved level of thermal comfort as a consequence of a diverse of triggers; for example, the relief that may occur a few hours after that flooding events ended taking place around different parts of the country. These specific events caused minor disruptions in terms of response and recovery in this case, and those increased levels of activity and positivity could be caused by multiple factors. Nevertheless, these periods of time when it is likely to spot moments and delays for certain generalised reactions to appear are significant and sensible aspects to consider in disaster phases, especially for events that could potentially cause major disruption.

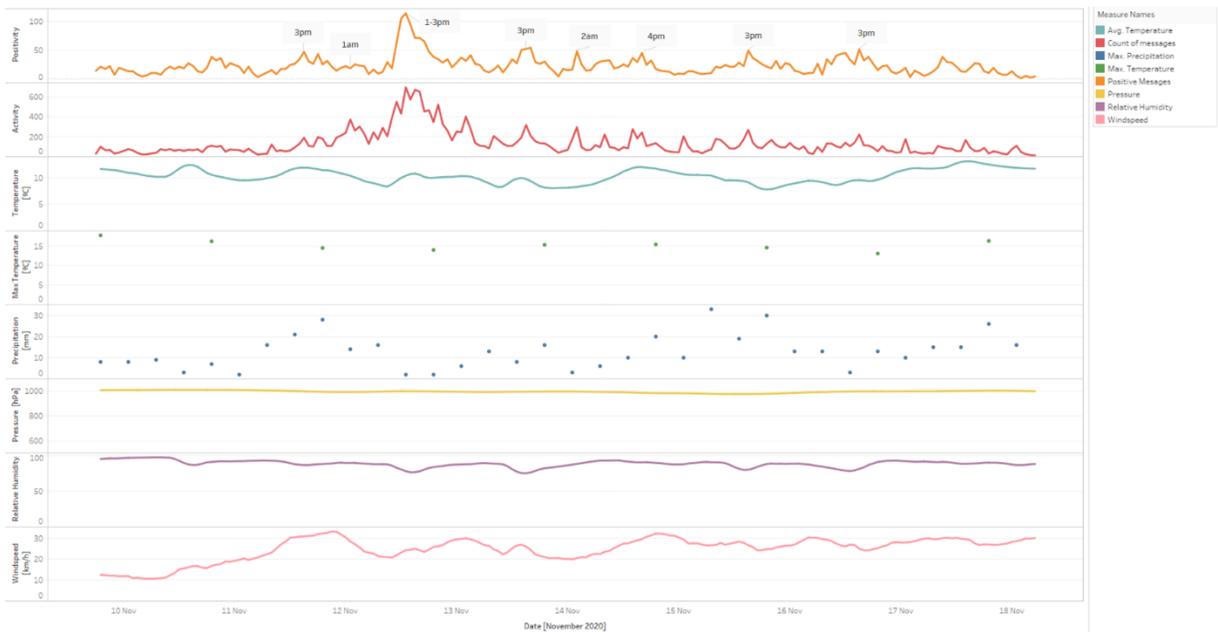

Figure 11: Levels of positive sentiment and user's activity along with climate variables on the UK Floods data collection during winter 2020.

Figure 12 shows a similar illustration as explained recently in Figure 11, in this case for the heatwaves collection. On the top, we plot the levels for each emotional indicators in alignment with user' activity and the set of climate variables. Another consideration is the division into three different episodes of heat waves during the periods summer 2020; the first during June-July, the second during July-August and the final heat wave during August-September. We can appreciate increased maximum temperatures between 23$^{rd}$ and 26$^{th}$ June of up to 33,4ºC, which appear to be linked with increased agitation and user's activity, as well as increased level of boredom. Potentially, this would indicate reduced thermal comfort levels caused by the heat wave. Once temperatures ceased, however, we notice another peak in user's activity on the 30$^{th}$ June. Similarly, the period the from the 30$^{th}$ July to 2$^{nd}$ August shows a similar phenomena with clearer correlations, plus we can also appreciate increased levels of increased anger with temperatures reaching up to 37,8ºC in few parts of the country on the 31$^{st}$ July. We also notice some other isolated peaks on emotions related to anger and boredom, despite these simply refer to a few outliers from repeated messages. In the last period, however, we can confirm again a clear correlation between user's activity, high temperatures, and some peaks of boredom and anger which are repeated at the end of this last period too. These correlations are especially noticeable between 10$^{th}$-12$^{th}$ August, 18$^{th}$-21$^{st}$ August, and also but in less degree between 7$^{th}$-9$^{th}$ and between 13$^{th}$-15$^{th}$ September.

In general, we expose the advantages of aligning behavioural indicators along with climate variables to provide with additional valuable information to be considered especially in different phases of a disaster, but also being applicable to episodes of extreme weather conditions to analyse general thermal comfort levels in a population from SMD inputs. Nonetheless, this experimental study shows the application in pilot study; influencing factors of thermal comfort are highly diverse and domain specific and therefore should require further evaluation.

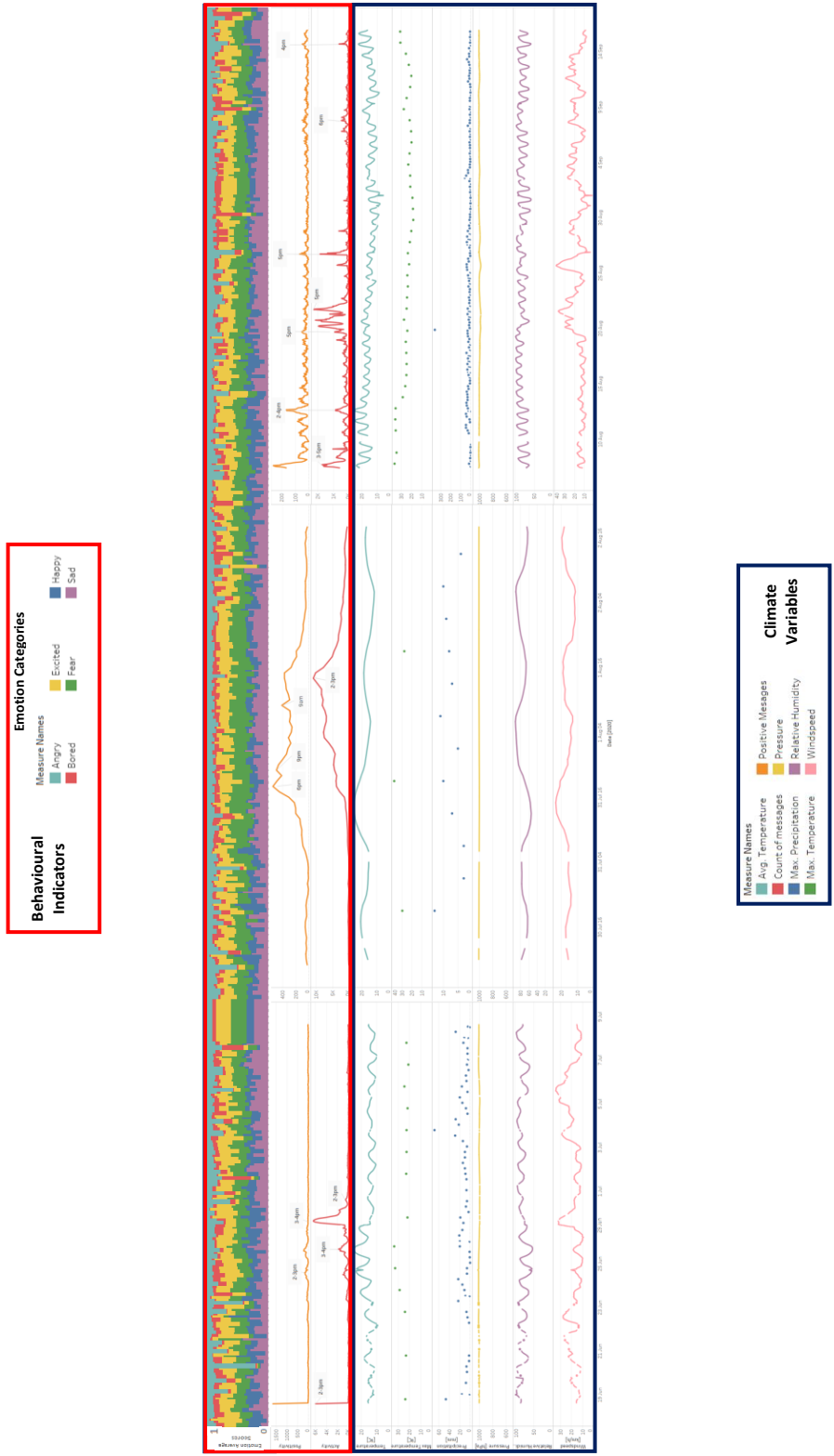

Figure 12: Levels of emotion and user's activity along with climate variables on the UK Heatwaves data collection in three different periods of time during summer 2020.

# 6 Conclusions

This research work focused on the qualitative analysis of social media data from disaster-related collections of floods and heatwaves. We use a methodology framework based on a combination of several artificial intelligence modules. The results of our analysis show novel insights in terms fusing approaches to 1) filter disaster-related messages using deep bidirectional Transformer models, 2) learn topic inference models and refine categorization through additional machine learning classifiers, 3) perform sentiment and behaviour analysis, and 4) match behavioural indicators with decoded climate variables from synoptical records to analyse thermal comfort. We also provide a semi-automatic method to discover key messages from user accounts to feedback the system.

It was shown that when we focus on the filtered messages, the positiveness of messages is around 8% for disaster, 1% for disaster and medical, 7% for disaster and humanitarian related messages. This low positivity suggests that many messages were coming from non-official sources, having plenty of negative connotations and barely telling anything valuable for the public. A trade-off between sources of relevant information coming both from governmental institutions and those from non-governmental institutions can complement each other to enrich public knowledge and informativeness. Also, informative messages should attach imagery to leverage the use of computer vision techniques for learning classification models from positive messages.

We highlight through our study the advantages of aligning behavioural indicators along with climate variables to provide with additional valuable information to be considered especially in different phases of a disaster, but also being applicable to extreme weather periods to analyse general thermal comfort levels in a population from SMD inputs.

Future work includes an analysis of geolocation inference methods for mapping the tweets according to their origin. On the side of behaviour indicators, a more detailed analysis of reactions would need to take into account likes, occurrences of forwarded messages (retweets), text analysis of responses to relevant messages as new potential indicators, among other factors of engagement predictions and powerful social determinants [39] to be included in the qualitative analysis. Finally, further validation strategies through quantified methods on the domain-specific influencing factors of thermal comfort will steer the use of behavioural indicators along with climate variables and provide better reliability through novel evaluations in the context of crisis and disaster response.

## Supporting Information

**S1 Supporting Information. List of unpublished work, technical reports, websites, and other electronic sources.** It contains 15 additional references.

## Acknowledgements

This research was funded by Belmont Forum's first disaster-focused funding Call Belmont Collaborative Research Action 2019: Disaster Risk, Reduction and Resilience (DR32019) which was supported by the Ministry of Science and Technology (MOST) of Chinese Taipei in partnership with funders from Brazil (FAPESP), Japan (JST), Qatar (QNRF), UK (UKRI), US (NSF), CNR (Italy). In particular, this research was funded by UKRI grant EP/V002945/1.